\documentclass[a4paper,twoside]{article}
\usepackage[authoryear]{natbib}
\bibpunct{(}{)}{;}{a}{}{,}
\usepackage{graphicx}

\date{} 
\title{\large\bf\flushleft An overview of the Laser Intererometer Space Antenna}
\author{\parbox{\textwidth}{\flushleft
\vspace{-0.5cm}
{\it D.A. Shaddock}\\
\vspace{0.4cm}
{\small \,Jet Propulsion Laboratory, California Institute of Technology, Pasadena CA 91109, USA}\\
{\small \,and Centre for Gravitational Physics, The Australian National University, Canberra ACT 0200, Australia}\\
{\small \,Email: Daniel.Shaddock@jpl.nasa.gov}}}
\begin{document}
\twocolumn[
\begin{changemargin}{.8cm}{.5cm}
\begin{minipage}{.9\textwidth}
\vspace{-1cm}
\maketitle
%
%
\small{{\bf Abstract:}\\The Laser Interferometer Space Antenna will detect gravitational waves with frequencies from 0.1~mHz to 1~Hz. This article provides a brief overview of LISA's science goals followed by a tutorial of the LISA measurement concept.}


\medskip
\medskip
\end{minipage}
\end{changemargin}
]
\small

\section{Introduction}

Einstein's Theory of General Relativity predicts that moving masses can produce propagating vibrations that travel through spacetime at the speed of light. These vibrations, referred to as gravitational waves (GWs), are produced by a quadrupolar acceleration of mass, in much the same way that electromagnetic radiation accompanies the acceleration of electric charges. The distortion of spacetime is transverse to the direction of wave propagation. GWs can be decomposed into two polarisations, corresponding to the axes of the stretching and squeezing, which are denoted by Ò+Ó (plus) and  Ò$\times$Ó (cross). We can sense GWs by monitoring changes in the distances between inertial proof masses. The gravitational wave amplitude is characterised by a dimensionless strain $h$, which determines the fractional amount of squeezing and stretching of spacetime.  

The direct detection of gravitational waves remains one of the most challenging areas of experimental physics.  Pulsar timing experiments \citep{Hobbs09} are searching for gravitational waves in the frequency band from 1-10~nHz.  A global network of laser interferometer gravitational wave detectors is now operational \citep{Whitcomb:CQG:08} targeting gravitational waves with frequencies from a few tens of Hertz up to a few kiloHertz with peak sensitivities near 100~Hz.  They are now operating at their design sensitivities, and are sensitive to gravitational wave strains as small as $10^{-21}$. Achieving this strain sensitivity requires measuring arm length changes of less than $10^{-18}$~m (over km scale baselines), a thousand times smaller than the diameter of a proton. This level of strain sensitivity is the result of decades of worldwide technology development, design, and commissioning. Future upgrades to these detectors to be carried out in the next few years will  improve the strain sensitivity by approximately an order of magnitude. Note that the volume of space that is probed increases as the cube of the strain sensitivity because interferometers respond directly to GW amplitude rather than GW power. Of equal importance, the lower frequency cutoff of the detectors will be reduced to around 10~Hz where sources are predicted to be more abundant. Even with substantially upgraded technology, however, future terrestrial detectors are expected to be limited to frequencies above a few Hertz due to gravity gradient noise \citep{Hughes:PRD:98}, fluctuating gravitational forces on the interferometer's proof masses. Candidate GW sources for the ground-based detectors include coalescing binary neutron stars, stellar core collapse triggering a Type II supernova, and rapidly rotating, non-axisymmetric neutron stars.

The Laser Interferometer Space Antenna (LISA) \citep{PPA} is a joint ESA-NASA project to operate the first dedicated gravitational wave detector in space. Like ground-based interferometers, LISA will sense gravitational waves by monitoring the changes in separation of inertial proof masses.  LISA is scheduled to launch around 2018 and will consist of three spacecraft forming an approximately equilateral triangle with 5 million kilometre arms.  Moving the detector to space not only allows the arms to be made very long, but it also eliminates the major disturbances to the proof masses that limit ground-based detectors at low frequencies. This opens up the rich region of the gravitational wave spectrum below 1~Hz. LISA will observe much larger objects, such as the inspirals and mergers of supermassive binary black holes, compact objects falling into supermassive black holes, as well as survey populations of binary compact objects (e.g. white dwarfs, neutron stars or stellar-mass black holes) in our galaxy.

\section{LISA Science}
GW observations will provide a unique window to the universe and offer a new set of capabilities compared to electromagnetic observations. The difficulty of detection of gravitational waves is a consequence of their weak interaction with matter. One benefit of the faintness of this interaction is that they are not attenuated or scattered on their way to the detector. This means that they can reveal information about areas or processes that are ordinarily obscured. GWs are emitted due to the bulk motion of massive objects. The wavelength of these waves is typically much larger than the source dimensions, which prevents us from imaging the source in the conventional sense. However, because the GW radiation is coherently emitted by the entire source, we can use both the magnitude and phase of the GWs to directly determine detailed information about the dynamics of the emitter.

There are a few important differences between LISA and an imaging electromagnetic observatory. First, LISA will be sensitive to sources throughout the sky. LISA can be viewed as three interferometers each with a quadrupolar antenna pattern which sweeps across the sky as the constellation's orbit evolves. This all-sky capability of LISA is of particular importance for detecting transient souces. Despite being an all-sky detector, the location of sources can be accurately determined.  The changing inclination of the detector provides amplitude modulation of gravitational waves while a slight frequency modulation results from Doppler shifts. These annual variations can provide around 10 arcminutes of source direction accuracy \citep{Hughes:06} depending on the source duration, signal to noise ratio and frequency.

LISA has a large frequency range spanning 4 decades from 0.1~mHz up to 1~Hz, with a peak sensitivity at around 3~mHz. The upper end of the frequency range is limited by the large size of the detector (the effect of GWs at high frequencies is averaged out over the 33 second round trip travel time of the light in the LISA arms). The low frequency limit arises from the difficulty of isolating the proof masses from disturbances. LISA will therefore complement the higher frequency observations of second generation ground-based interferometers and will be sensitive to different types of gravitational wave sources. 

One of LISA's primary science goals is to trace the growth and merger history of massive black holes and their host galaxies. Binary black holes with $10^5 - 10^7$ solar masses will coalesce within LISA's frequency band.  LISA will be able to track these mergers at large redshifts and will determine the parameters of the black holes with high precision. 

Another important LISA science goal involving black holes is the observation of compact objects falling into supermassive black holes. These extreme mass-ratio capture binaries will allow LISA to perform high precision mapping of the spacetimes around massive objects in galactic nuclei and provide quantitative observational tests of General Relativity. 

The most numerous sources in the LISA band are expected to be ultra-compact binaries in our galaxy. LISA will survey the populations of these stellar-mass binaries, mostly white-dwarf binaries, neutron stars or black holes. This study will reveal information about the structure of the galaxy and provide insight into dynamics in galactic nuclei.  LISA is expected to  detect so many galactic binary inspirals that it will be ``confusion noise'' limited at some frequencies, meaning that there will be so many signals detected that it may be impossible to distinguish between individual sources. Approximately 10 known binaries with large signal to noise ratios and well known source parameters will be used as verification binaries \citep{Nelemans:2006AIP}. Failure to detect these signals would indicate serious problems with our understanding of General Relativity. For an in-depth discussion of LISA science see references \citep{Hughes:06, LISAScienceCase}.

\section{LISA Mission Overview}\label{Overview}

In space, the interferometer arms can be made very long, around a million times longer than ground-based interferometers. LISA will have arm lengths of five million kilometers. Longer arms amplify the effect of low frequency gravitational waves, allowing LISA to relax the displacement sensitivity by approximately 6 orders of magnitude compared to ground-based interferometers, yet maintain an approximately equivalent strain sensitivity. The LISA constellation is in an Earth trailing, heliocentric orbit approximately $20^\circ$ behind Earth. The LISA constellation appears to cartwheel around the sun with an inclination angle of $60^\circ$. This configuration was chosen to passively maintain the approximately equilateral triangle formation without any orbital adjustments. The orbits also produce a constant solar illumination which results in a benign thermal environment.

\begin{figure}[ht]
\includegraphics[width = 0.5\textwidth]{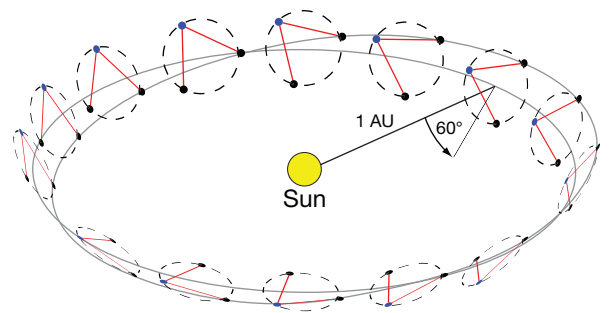}
\caption{Illustration of the LISA orbit showing the 60$^\circ$ angle of inclination of the LISA constellation plane.  The LISA constellation trails the Earth by 20$^\circ$. \label{LISAOrbit}}
\end{figure}

Going to space has many advantages and enables several simplifications compared to ground-based interferometers. For example, the micro-gravity environment is naturally suited to freely floating masses, eliminating the need for complex seismic isolation systems. The abundance of free vacuum in space allows very long arm lengths and a simple interferometer topology, a very long Michelson interferometer. Ground-based detectors employ more complex inteferometer topologies to maximize the signal response for a given arm length.  Some things, however, are more complicated in space. One complication is that the LISA arm lengths change by tens of thousands of kilometers over a period of a year, due mainly to perturbations by the Earth's gravity. There is no way to anchor the spacecraft, so the arm lengths cannot be easily fixed as they can on Earth. These long term length fluctuations are the root of several technical problems. For example, the angles between the outgoing beams must be adjusted to ensure that the laser light is exchanged correctly between spacecraft. The relative spacecraft motion also Doppler shifts the laser frequency and places challenging requirements on the interferometric readout. Finally, the mismatched arm lengths couple laser frequency noise into the GW readout.

Gravitational waves are observed by their effect on the motion of objects. Broken down to the simplest elements, we need objects to measure and a way to measure their displacement. The objects to be measured are referred to as proof masses. Like ground-based detectors, the displacement of these objects is measured using laser interferometry. LISA's sensitivity will be limited by proof mass acceleration noise in the low frequency region of its detection band, and by the interferometer measurement noise, dominated by shot noise, at high frequencies.

\subsection{Proof masses and the disturbance reduction system}
Each LISA spacecraft will house two proof masses: 46 mm gold-platinum alloy cubical masses with a gold coating. The challenge comes in isolating these proof masses from unwanted forces. The LISA measurement calls for a proof mass acceleration noise of $\sim10^{-15}$~m/s$^2$/$\sqrt{\rm Hz}$ near 1~mHz. To achieve this the spacecraft must shield the proof masses from the solar wind and the entire spacecraft is designed with gravitational balancing in mind \citep{Armano:CQG:05}. The proof mass material is tailored to minimize its magnetic interaction, and the magnetic environment of the spacecraft and internal components will be carefully controlled \citep{Hueller:CGQ:05}.

Each proof mass will float inside its housing with no mechanical contact \citep{LTP}. A combination of electrostatic sensors and optical metrology will continuously monitor the position and orientation of each mass with respect to the housing. The proof mass is part of a capacitive bridge and is voltage biased via its capacitive coupling to a set of injection electrodes. Motion sensing will be performed by differentially reading the capacitance between proof mass and electrode pairs on opposite faces. By using 6 electrode pairs and 6 capacitive bridges, all 6 degrees of freedom can be read out. Capacitive actuators will suspend the proof masses in the non-sensitive direction to prevent the masses from colliding with the housing walls. During science operations the spacecraft will follow the proof mass in the sensitive direction. For drag-free control, the proof mass to spacecraft displacement readout sensitivity must be better than 2~nm$/\sqrt{\rm {Hz}}$ in the sensitive direction. Sensitivities may be relaxed by a factor 10 at low frequencies along all other degrees of freedom as the requirements are intended only to reduce crosstalk due to imperfections. 

The proof masses and associated hardware will be flown on the LISA precursor mission, LISA Pathfinder (LPF)  \citep{LTP}, which is currently scheduled to launch no sooner than the end of 2009.  LPF is nominally identical in all its components to those which will be flown on LISA, and will enable testing of these components in a LISA-like environment. LPF will orbit the L1 point, located 1.5 million kilometres from the Earth in the direction of the Sun.

Low noise, micro-Newton thrusters allow the spacecraft to track proof mass motion in the sensitive direction. The thrusters primarily push back against the solar irradiance which has a DC acceleration value of $10^{-7}$~m/s$^2/\sqrt{\rm Hz}$. The thrusters are also responsible for maintaining the spacecraft orientation based on feedback sensors from both the proof mass and the interferometer's wavefront sensing.  Thrusters for the LISA Test Package on LPF are based on Field Emission Electric Propulsion \citep{FEEP}. NASA have developed colloid micro-Newton thrusters \citep{Ziemer} that will also be tested on LPF.
 
\subsection{LISA Interferometry}\label{Inteferometry}
The LISA displacement measurement is based on heterodyne interferometry \citep{Bobroff}. In heterodyne interferometry, two lasers with different frequencies are interfered to produce a beat note at their frequency difference. Displacement is detected by measuring the phase of this beat note, where one optical wavelength of path length change would produce a phase shift of one cycle in the beat note. The displacement readout is therefore self-calibrating down to the knowledge of the laser wavelength. A heterodyne interferometer has no preferred ``lock point'' and can track the phase over millions of cycles. The Doppler shifts cause the LISA beat note frequency to vary from 2~MHz up to nearly 20~MHz over the course of a year. The GW signal appears as mHz phase modulation sidebands on this MHz signal.

The change in separation of proof masses along each interferometer arm is determined by combining three measurements \citep{HeinzelCQG06}; two local measurements of the proof masses' displacement with respect to their optical benches, and one interspacecraft measurement of the relative displacement of optical benches. The local metrology is practically identical to the interferometry used on LISA Pathfinder \citep{RobertsonCQG05}. The interspacecraft interferometry is based on the same heterodyne interferometry but has a few additional complications. The optical path includes the 40~cm diameter telescopes needed to transmit and receive the light between spacecraft. The outgoing beam has a 3~dB half cone divergence of 1.6~$\mu $rad \citep{Astrium} which results in a beam diameter of 16~km at the far spacecraft. After accounting for losses in the optical chain, only $\sim$100~pW of light is available for the measurement. This power level has a shot noise limited measurement error of 8~pm/$\sqrt{\rm{Hz}}$. If this light were simply reflected back to the distant spacecraft then the detected signal would be completely overwhelmed by shot noise. Instead an optical transponder arrangement is used, where another laser is phase-locked to the low power received beam and then transmitted back to the originating spacecraft.  

The weak light returning from each arm is not interfered directly on a beamsplitter as in a conventional Michelson interferometer. Instead, each beam is interfered with a bright local oscillator derived from the outgoing beam. The differential displacement of the two arms, the Michelson observable, is then constructed by electronically recombining the phase measurements from each arm. In this way a Michelson interferometer configuration can be synthesized from the combination of four measurements (two at the ends of the interferometer arms and two of the return beams at the originating spacecraft). The total shot noise from all four measurements yields a round trip error of 16~pm/$\sqrt{\rm{Hz}}$ and is the dominant contribution to the LISA measurement noise at high frequencies.

\subsection{Laser frequency noise suppression}\label{FrequencyNoise}

Laser interferometers use the wavelength of the laser light as the yardstick for measuring displacement.  A change in the laser wavelength (or equivalently the laser frequency) will be misinterpreted as a change in measured separation of the proof masses. For gravitational wave detection, we can take advantage of the quadrupole nature of gravitational waves and measure the difference between two equal length arms. In this scenario  the effect of laser frequency noise cancels when the light from the arms is recombined.  Alternatively, locking the laser to a stable length reference, such as an optical cavity, can stabilise the frequency.  LISA uses a novel combination of these standard techniques. 

LISA laser frequency stabilisation will be performed in two stages, similar to the two-stage process used in ground-based gravitational wave detectors.  First, the lasers are pre-stabilised to a reference optical cavity. In principle it is possible to transfer the fractional length stability of the cavity to the fractional frequency stability of the laser frequency. Researchers at the Goddard Space Flight Center constructed an optical cavity from ultra-low expansion (ULE) glass, and placed it in a stable, LISA-like thermal environment. Their results demonstrate that it is possible to achieve around $10^{-13} /\sqrt{\rm{Hz}}$ laser frequency stability across the LISA band \citep{Mueller}.

The second stage of stabilisation locks the laser frequency to the 5 million km LISA arms.  This Òarm-lockingÓ will improve laser frequency stability by an additional factor of 100 beyond the cavity pre-stabilised noise level. Any laser frequency locking system is limited by the quality of the reference. In the 5 million km LISA arms we have custom made length references that are uniquely stable throughout the LISA band. The challenge in stabilising to the arms, a technique referred to as arm-locking, is to build a high gain feedback control system in the presence of the large 33.3 second round trip delay. Ordinarily, such a delay would necessitate reducing the bandwidth and gain of the feedback until a stable system is achieved. In LISA, this would limit the unity gain bandwidth to less than $30$~mHz and provide only a small amount of noise suppression at the lower frequencies of the LISA band. Researchers at the Australian National University first noted \citep{Sheard:arm_locking:03} that the bandwidth restriction can be circumvented by using a rather unconventional feedback controller. There have since been several simulated \citep{Sylvestre:simluation:02} and experimental \citep{Marin,Sheard:06,Thorpe} validations of the control system stability and noise suppression. More recently, the arm-locking control system was expanded to include signals from both arms, resulting in substantial improvement in laser frequency noise \citep{herz:freq_stabil:05,Sutton}.

\begin{center}
\begin{figure}[!h]
\includegraphics[width = 0.5\textwidth]{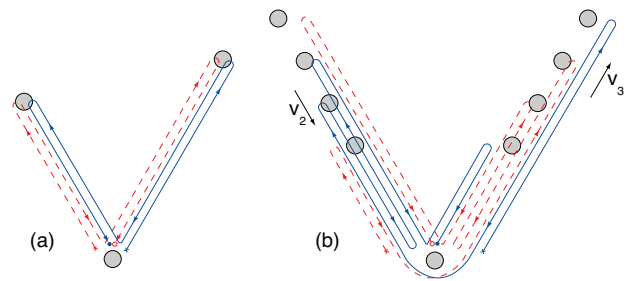}
\caption{Interferometer configurations corresponding to (a) length correcting and (b) velocity correcting Michelson time delay interferometry combinations. The LISA spacecraft are indicated by solid circles. \label{Michelson}}
\end{figure}
\end{center}

\begin{figure}[ht]
\includegraphics[width = 0.5\textwidth]{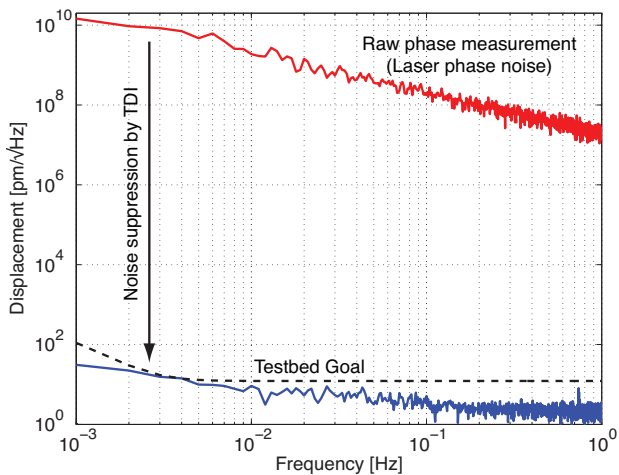}
\caption{Demonstration of laser frequency noise suppression by a factor of $\sim10^9$ using a simplified version of TDI. \label{Alpha}}
\end{figure}

The remaining laser frequency noise will be removed in post-processing using a technique called time delay interferometry (TDI) \citep{TintoPRD99,tinto:TDI:03}. TDI forms linear combinations of the phase measurements that are free from laser frequency noise. TDI exploits the fact that the laser frequency noise enters each measurement with a well known transfer function that depends only on the light travel time of the LISA arms.  TDI can be understood as a way to synthesize an interferometer that has equal arm lengths. Figure \ref{Michelson} shows combinations that correct for mismatched (a) arm length and (b) velocity \citep{ShaddockPRD03}. This is achieved by time delaying the data streams using high performance interpolation algorithms. These same linear combinations maintain the gravitational wave signal, as it is contained in the difference of the arms while the laser noise is common to the arms. Frequency noise cancellation was first experimentally demonstrated in 2006 in the LISA interferometry test bed at the Jet Propulsion Laboratory using a simplified version of TDI. The measured noise spectra presented in Figure \ref{Alpha} show that laser frequency noise is suppressed by up to 9 orders of magnitude, down to the intrinsic noise of the test bed. In 2008 the test bed was upgraded to include the complete implementation of TDI, including interpolation and noise from the ultra-stable oscillators used in the phase measurement.

\section{Australia's role in LISA}

Australia has a strong track record in the field of interferometric gravitational wave detection. Groups operating at The Australian National University, the University of Western Australia, the University of Adelaide and elsewhere have active research programs to develop technology and data analysis techniques for ground-based detectors.  As a result of these contributions, Australian researchers now have access to data from the network of ground-based gravitational wave detectors. Although Australia has no official involvement in the LISA project, there have been several important technical contributions from the ANU, and there are many Australians now working on the LISA project in both Europe and the US.  It remains to be seen if there is sufficient scientific and political will for Australia to pursue a more substantial role in LISA and guarantee Australian scientists access to LISA data.

As the vast majority of our contact with the rest of the universe is through electromagnetic radiation, it is difficult to predict everything that this new form of astronomy will deliver to us. One thing that can be learnt from experiences with other forms of astronomy is that gravitational wave astronomy may one day begin to discover entirely new and unexpected sources. This alone is perhaps the most compelling reason to pursue the detection of gravitational waves. 

\section{Acknowledgments}
I would like to thank the Astronomical Society of Australia for the invitation to submit this paper as a result of an invited talk at the ASM. The work reported on here is a summary of the work of the LISA community and I gratefully acknowledge the hard work of many researchers around the world.  Parts of this research were performed at the Jet Propulsion Laboratory, California Institute of Technology, under contract with the National Aeronautics and Space Administration. This research was also supported by the Australian Research Council's Discovery Projects funding scheme (project number DP0666437). 

\bibliography{bibliography}

\end{document}